\documentclass[letter]{jpconf}
\usepackage{graphicx} 
\usepackage{amssymb} 
\usepackage{fixmath} 
\usepackage{bm} 

\newcommand{ \be }{\begin{eqnarray}}
\newcommand{ \ee }{\end{eqnarray}}
\newcommand{ \la }{\langle}
\newcommand{ \ra }{\rangle}

\newcommand{ \ds }{\displaystyle}

\newcommand{ \mean }[1]{\left\langle #1 \right\rangle}   

\def\snn{$\sqrt{s_{NN}}$}

\def\P{$\cal P$}
\def\CP{$\cal CP$}

\newcommand{ \psirp }{\Psi_{RP}}
\newcommand{ \phia }{\phi_{\alpha}}
\newcommand{ \phib }{\phi_{\beta}}

\usepackage{color}

\begin{document} 

\title{Local strong parity violation and new possibilities in
  experimental study of non-perturbative QCD }

\author{Sergei A. Voloshin}

\address{Department of Physics and Astronomy,
Wayne State University, Michigan 48201, USA }

\ead{voloshin@wayne.edu}

\begin{abstract}
Quark interaction with topologically non-trivial gluonic fields,
instantons and sphalerons, violates \P~ and \CP~ symmetry.
In the strong magnetic field of a non-central nuclear collision
such interactions lead to the charge separation along the magnetic
field, the so called chiral magnetic effect, which manifests local
parity violations.
An experimental observation of the chiral magnetic effect would 
be a direct proof for the existence of such physics.
Recent STAR results on charge and the reaction plane dependent 
correlations are consistent with theoretical expectations for the
chiral magnetic effect.
In this paper I discuss other approaches to
experimental study of the local parity violation, and propose 
future measurements which can clarify the picture. 
In particular I propose to use central body-body U+U collisions 
to disentangle correlations due to chiral magnetic effect from
possible background correlations due to elliptic flow.
Further more quantitative studies can be performed with
collision of isobaric beams.
\end{abstract}

 \section{Introduction. Local parity violation.}
It is widely accepted that Quantum Chromodynamics (QCD) 
is the theory of strong interactions. 
Perturbative QCD is firmly established and thoroughly tested experimentally. 
In the non-perturbative sector, QCD links chiral symmetry breaking and the 
origin of hadron masses to the existence of topologically non-trivial 
classical classical gluonic fields 
describing the transitions between the vacuum states with 
different Chern-Simons numbers.  
Quark interactions with such fields change the quark helicity 
and are $\cal P$~and $\cal CP$~odd. 
For a review, see~\cite{Shuryak:1997vd,Diakonov:2002fq}.
It was first suggested in~\cite{Kharzeev:1998kz} 
to look for such metastable $\cal P$~and $\cal CP$~odd domains in 
ultra-relativistic heavy ion collisions. 
The possibilities for an experimental detection of this  
{\em  local strong  parity  violation} was discussed
in~\cite{Kharzeev:1998kz,Voloshin:2000xf,Finch:2001hs}.

Originally~\cite{Kharzeev:1998kz}, it was proposed to look for the effect 
by detecting the non-statistical fluctuations in the variable
\be
J=\sum_{\pi^+, \pi^-} \frac{\ds (\vec{p}_{\pi^+} \times \vec{p}_{\pi^-})_z}
{\ds p_{\pi^+} p_{\pi^-}}.
\ee 
In~\cite{Voloshin:2000xf} it was shown that $J$ is directly related
to the difference in the event planes reconstructed from positive 
and negative  particles.
This difference is similar to the effect of a (electro) magnetic field
oriented along the beam direction (parallel or anti parallel).
The existence of such a field in a symmetric collision would
constitute parity violation. 
Experimental measurements of the effect by NA49 Collaboration
revealed a signal consistent with zero~\cite{volo99}.
Unfortunately, this observable have not been measured at RHIC yet.
 
More recently, it was noticed~\cite{Kharzeev:2004ey,Kharzeev:2007tn} 
that in non-central nuclear collisions such domains 
can demonstrate themselves via the asymmetry in the emission 
of positively or negatively charged particle perpendicular to the reaction plane.
Such charge separation is a consequence of the difference in the number
of particles with positive and negative helicities positioned
in the strong magnetic field ($\sim 10^{15}$~T) 
of  a non-central nuclear collision, the so-called 
{\em chiral magnetic effect}~\cite{Kharzeev:2004ey,Kharzeev:2007jp}. 
The same phenomenon can also be described in terms of the induced electric
field that is parallel to the static external magnetic field, 
which occurs in the presence of topologically non-trivial vacuum 
solutions~\cite{Fukushima:2008xe}.
The direction of the separation varies event by 
event in accord with the sign of the topological charge of the domain,
and the observation of the effect is possible only by correlation techniques.
An observable directly sensitive to the charge
separation effect, has been proposed in~\cite{Voloshin:2004vk}.
It is discussed in more detail below.

According to Refs.~\cite{Kharzeev:2004ey,Kharzeev:2007tn,Kharzeev:2007jp}
the charge separation could lead to asymmetry in particle production
 $\sim Q/N_{\pi^+}$, where $Q=0,\pm 1,\pm 2, ...$ 
is the topological charge and
 $N_{\pi^+}$ is the positive pion multiplicity in 
one unit of rapidity 
-- the typical scale of such correlations.
The charge separation effect is expected to depend strongly on
deconfinement
and chiral symmetry restoration~\cite{Kharzeev:2007jp}, 
and the signal might be greatly suppressed or completely absent
at an energy below that at which a quark-gluon plasma can be formed.

\section{Chiral magnetic effect and charge dependent azimuthal correlations}

\begin{figure}
\begin{minipage}[t]{0.5\textwidth}
\includegraphics[width=1.05\textwidth]{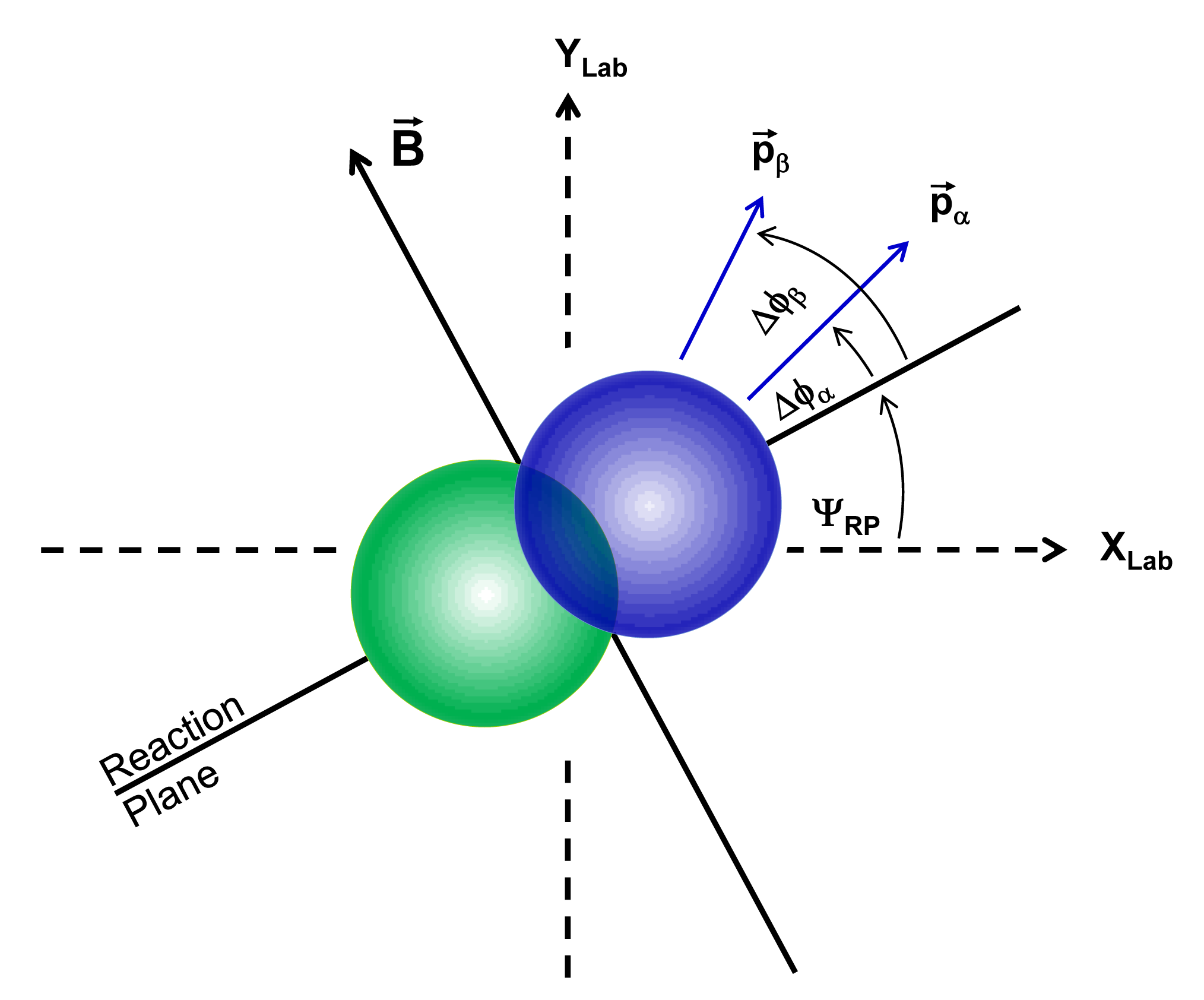}
\end{minipage}
\hspace{0.0\textwidth}
\begin{minipage}[b]{0.45\textwidth}
 \caption{Schematic view of non-central nuclear collision showing the
   definition of angles in 
Eqs.~\ref{eq:expansion},~\ref{eq:correlator}.
}
 \label{fig1}
\end{minipage}
\end{figure}

In non-central nuclear collisions particle distribution in azimuthal
angle is not uniform. The deviation from a flat distribution is
called anisotropic flow and often is described by the Fourier
decomposition~\cite{Voloshin:1994mz} (for a review,
see~\cite{Voloshin:2008dg}):
\be
 \frac{dN_\alpha}{d\phi} &\propto&  1+ 2 v_{1,\alpha} \cos(\Delta \phi)+
2\, v_{2,\alpha} \cos(2\Delta\phi)+... 
\nonumber \\
&+& 2  a_{1,\alpha} \sin(\Delta \phi) +
2\, a_{2,\alpha} \sin(2 \Delta \phi) +... \, ,
\label{eq:expansion}
\ee
where $\Delta \phi =(\phi-\psirp)$ is the particle azimuth 
relative to the reaction plane (see Fig.~1),
$v_1$ and $v_2$  account for directed and elliptic flow. 
Subscript $\alpha$ is used to denote the particle type.
Due to the ``up-down'' symmetry of the collisions $a_{n}$ coefficients 
are usually omitted.
Chiral magnetic effect violates such a symmetry. 
Although the  ``direction'' of
the violation fluctuates event to event and on average is zero,
in events with a particular sign of the topological charge, the
average is not zero. 
As a result, it leads to a non-zero contribution to correlations,
e.g. $\mean{a_{n,\alpha} a_{n,\beta}}$,
where $\alpha$ and $\beta$ denote the particle type.
One expects that the first harmonic would account for the most of the
effect.
Below, if not explicitly mentioned, it is assumed that $n=1$. 
Note that only particles originated from (interacted with) the same domain 
are correlated. The size of the domain is expected to be less than
1~fm, which means that the correlated particles are likely to be
within about one unit of rapidity from each other.
To measure $\mean{a_\alpha a_\beta}$, 
it was proposed~\cite{Voloshin:2004vk} to use the correlator:
\be
\hspace*{-2cm}
& \mean{ \cos(\phia +\phib -2\psirp) } = 
\label{eq:obs1}
\mean{\cos\Delta \phia\, \cos\Delta \phib} 
-\mean{\sin\Delta \phia\,\sin\Delta \phib}
\label{eq:cossin}
\\ 
& =
[\mean{v_{1,\alpha}v_{1,\beta}} + B_{in}] - [\mean{a_\alpha a_\beta}
+ B_{out}] \approx - \mean{a_\alpha a_\beta} + [ B_{in} - B_{out}].
\label{eq:correlator}
\ee
This correlator represents the
difference between correlations ``projected'' onto 
the reaction plane and the correlations projected onto an axis
perpendicular to the reaction plane. 
The key advantage of using such a difference is that it
removes all the correlations among
particles $\alpha$ and $\beta$ that are not related to the reaction plane 
orientation.
The contribution given by the term $\mean{v_{1,\alpha}v_{1,\beta}}$
can be neglected because directed flow averages to zero in
a rapidity region symmetric with respect to mid-rapidity.

{\bf \textsl{RP dependent, ``physics'', background.}}
The remaining background in Eq.~\ref{eq:correlator},  $B_{in}
-B_{out}$, are due to processes in which particles $\alpha$ and
$\beta$ are products of a cluster (e.g. resonance, jet,
 di-jets) decay, and the cluster itself exhibits elliptic
flow or decays (fragments)
differently when emitted in-plane compared to out-of-plane.
The corresponding contribution to the correlator can be estimated as:
\be  
\hspace{-1cm}
\la \cos(\phi_\alpha + \phi_\beta -2\psirp) \ra =
\frac{ N_{\frac{clust}{event}} N_{\frac{pairs}{clust}}} 
{ N_{\frac{pairs}{event}}}    
          \,
\la \cos(\phi_\alpha + \phi_\beta -2\phi_{clust}) \ra_{clust}
\; v_{2,clust},
\label{eq:resonance}
\ee 
where $\la ... \ra_{clust}$ indicates that the average is performed only
over pairs consisting of two daughters from the same cluster.
This kind of background can not be easily suppressed. To address its
contribution one has to rely on model calculations or perform
experiments where the relative contribution of chiral magnetic effect
and background can be varied, see Section~\ref{sec:future}.

{\bf \textsl{RP independent background.}}
The reaction plane is not known experimentally, and has to be
estimated event-by event. 
Note that the second order reaction plane~\cite{Poskanzer:1998yz}, 
reconstructed from strong elliptic flow~\cite{Ackermann:2000tr}, 
is sufficient for this measurement.
Usually, the correlator Eq.~\ref{eq:correlator} is evaluated with the
help of 3-particle correlations:
\be
\la \cos(\phi_\alpha + \phi_\beta -2\phi_c) \ra \approx
\la \cos(\phi_\alpha + \phi_\beta -2\psirp) \ra v_{2,c}.
\ee
This factorization can be broken, e.g. by correlations from a cluster
decay to all three, $\alpha$, $\beta$, and $c$ particles.
Such contribution can be greatly suppressed by appropriate choice of
particle $c$, see~\cite{GangHere}.

\section{STAR results.}

\begin{figure}
\begin{minipage}[t]{0.5\textwidth}
\includegraphics[width=1.1\textwidth]{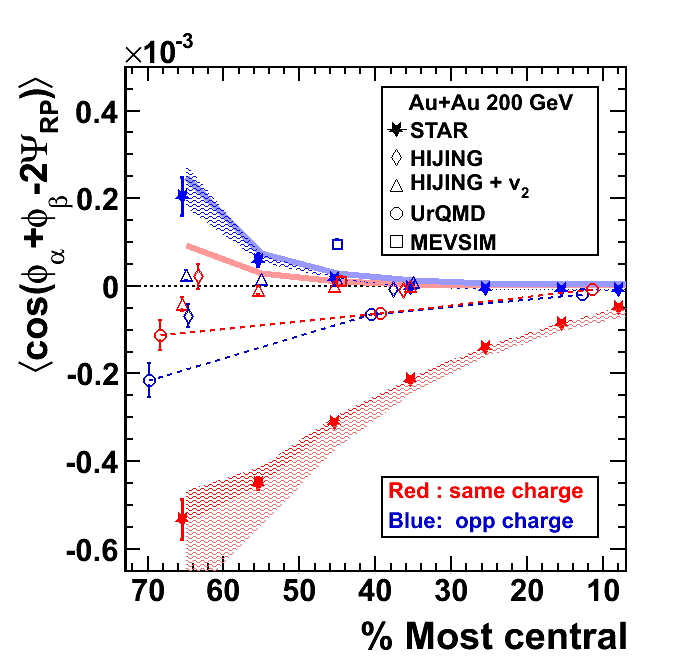}
\end{minipage}
\hspace{0.05\textwidth}
\begin{minipage}[b]{0.45\textwidth}
 \caption{(Taken from~\cite{:2009txa}) 
  STAR results compared to simulations for 200~GeV Au+Au.
  Blue symbols mark  {opposite-charge}
  correlations, and red are  {same-charge}.  
  The shaded  bands show the systematic error due to uncertainty 
  in $v_2$ measurements.  
  In simulations the true 
  reaction plane from the generated event was used.
  Thick solid lighter colored lines represent non reaction-plane
 dependent contribution as estimated by HIJING. 
 Corresponding estimates from UrQMD are about factor of two smaller.
\vspace*{3mm}
}
 \label{figSTAR}
\end{minipage}
\end{figure}

Recently, the charge and reaction plane dependent correlations
for Au+Au and Cu+Cu collisions at \snn=200~GeV and \snn=62~GeV
have been published by the STAR Collaboration~\cite{:2009txa,:2009uh}. 
The correlations are reported for charged particle in the region
 $|\eta| < 1.0$ with $p_t > 0.15$~GeV/c. 
Figure~\ref{figSTAR} shows STAR results for Au+Au collisions at
\snn=200~GeV compared to predictions from different event generators. 
Note that the latter are not zero, and is due to discussed above
reaction plane dependent background. 
To illustrate this better, the results from UrQMD event generator 
are connected by dashed lines.
Reaction plane independent background, for a particular method
employed in this analysis, is shown by thick lines.
Opposite-charge correlations are smaller than same-charge
correlations. 
This observation led to the proposal~\cite{Kharzeev:2007jp} 
that back-to-back
correlations may be suppressed due to the opacity of the medium.
The correlations are weaker in more central collisions compared to more
peripheral collisions, which partially can 
be attributed to dilution of correlations
which occurs in the case of particle production from multiple sources. 
To compensate for the dilution effect, in particular
when comparing Au+Au results to Cu+Cu, STAR also presented
results multiplied by the number of participants 
(for a plot, see~\cite{:2009txa}). 
There, the same and opposite sign correlations
clearly exhibit very different behavior. 
The opposite sign correlations in Au+Au and Cu+Cu collisions are
found to be very close at similar values of $N_{part}$ in rough
qualitative agreement with the picture in which their values are mostly
determined by the suppression of back-to-back correlations.
The difference in magnitude between same and opposite sign correlations
is considerably smaller in Cu+Cu than in Au+Au,
qualitatively in agreement with the scenario of stronger suppression 
of the back-to-back correlations in Au+Au collisions.

\begin{figure}
\begin{minipage}[t]{0.5\textwidth}
\includegraphics[width=1.03\textwidth]{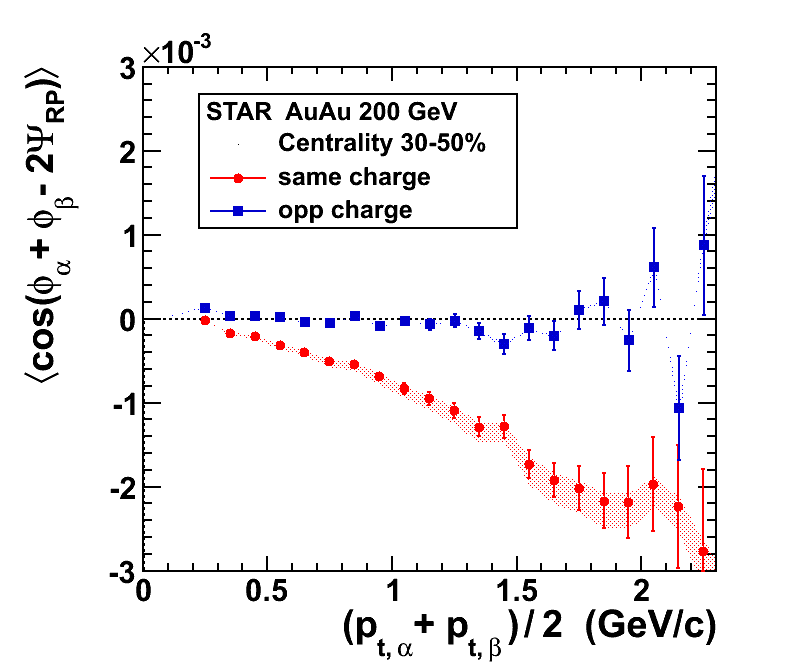} 
	\centerline{(a) }
\end{minipage}
\hspace{0.0\textwidth}
\begin{minipage}[t]{0.5\textwidth}
\includegraphics[width=1.03\textwidth]{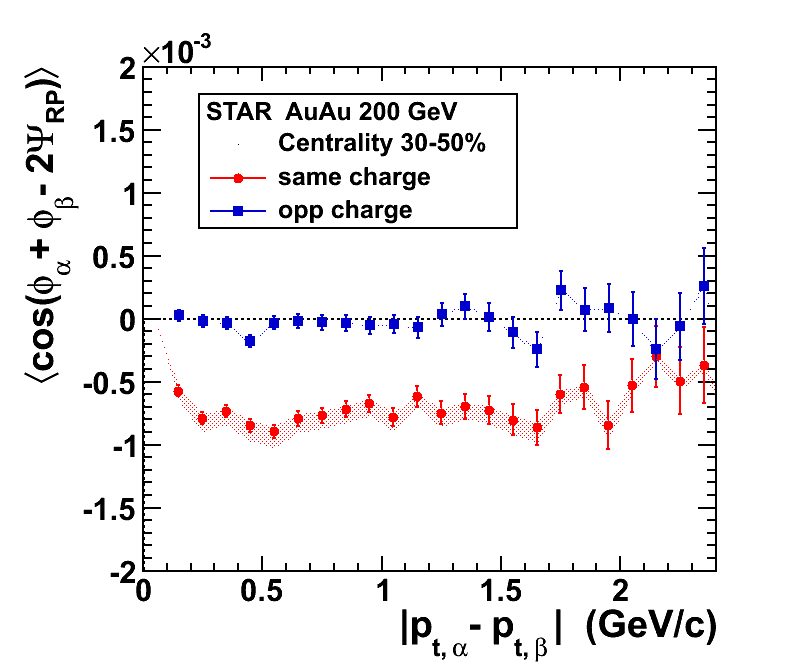} 
	\centerline{(b) }
\end{minipage}
 \caption{(Taken from~\cite{:2009txa,:2009uh}). 
Au+Au at 200 GeV~\cite{:2009txa}. The correlations dependence 
(a) on sum  and (b) the difference
of the magnitude of transverse momenta.
} 
\label{figPT}
\end{figure}

\begin{figure}
\begin{minipage}[t]{0.55\textwidth}
\includegraphics[width=1.\textwidth]{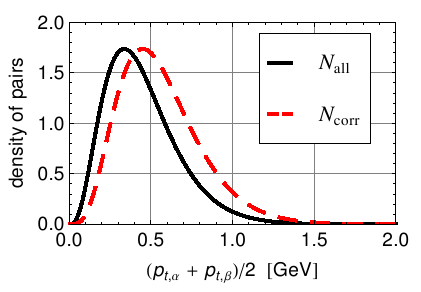}
\end{minipage}
\hspace{0.0\textwidth}
\begin{minipage}[b]{0.42\textwidth}
 \caption{Taken from~\cite{Bzdak:2009fc}. 
Transverse momentum distribution of pairs from clusters, shown in red,
compared to distribution of all pairs. 
\vspace{3mm}
}
 \label{figKoch}
\end{minipage}
\end{figure}

Somewhat unexpected result was the dependence of the signal on 
the transverse momentum of the two particles, see Fig.~\ref{figPT}.
It was found that the signal is not concentrated in the low $p_t$
region as naively might be expected for \P-violation
(non-perturbative) effects, and
that the correlation depends very weakly on
$|p_{t,\alpha}-p_{t,\beta}|$. 
An interesting possible explanation for such a dependence was found 
in~\cite{Bzdak:2009fc}. 
It was found that 
such a dependence can be naturally understood if one assumes 
that the $p_t$ distribution of correlated pairs has somewhat harder 
spectrum compared to that of all (random) pairs.
The ``trick'' is that even if the distribution of pairs from clusters is
only slightly harder, the relative weight of correlated pairs
increases with $p_t$.
Figure~\ref{figKoch} shows the pair distributions in transverse momentum
that was obtained in~\cite{Bzdak:2009fc} using STAR data as an input. 

\section{Future program.}  
\label{sec:future}
The STAR observation of the charge
dependent azimuthal correlations consistent with the theoretical
expectations for the chiral magnetic effect can be a beginning of an
exciting program.
Of course, one would need first to confirm that the observed correlations indeed
are related to the local parity violation. 
If confirmed, it will open the door for 
{\em direct} experimental measurements of the effects 
from non-perturbative sector of QCD, so far not possible.

\subsection{\bf \textsl{Experiment}.}
Below I discuss several directions for a future experimental program.
In particular I propose to use central body-body U+U collisions 
to test if the observed correlations are indeed related to the
strong magnetic field, as they must be in chiral magnetic effect. 
More detailed study of the dependence of the effect on the magnetic
field can be achieved with collisions of isobaric nuclei.   
Measurements aimed on understanding the properties of the clusters 
in multiparticle production would be another important part of
the program.
\begin{itemize}
\item {\sl Central body-body U+U collisions.} 
Ambiguity in the interpretation of the STAR results lies 
in the difficulty to eliminate/suppress the RP dependent
background, $B_{in}-B_{out}$, which by itself originates in
the anisotropy of particle production relative to the reaction
plane -- elliptic flow. To eliminate/suppress elliptic flow, and at the
same time preserve strong magnetic field needed for the chiral
magnetic effect does not look realistic. But the opposite is possible, 
one can have collisions with strong elliptic flow and no (or
almost no) magnetic field. 
This can be achieved in central body-body U+U collisions.  
Uranium nuclei are not spherical and have roughly ellipsoidal shape.
Central collision, when most of the nucleons interact, can have
different geometry, ranging from the so called tip-tip collisions to
body-body collisions~\cite{Nepali:2007an}, see Fig.~\ref{figUU}. 
Unlike tip-tip collisions, body-body ones would exhibit strong elliptic flow. 
Neither would lead to a strong magnetic field; consequently, 
a very weak signal due to chiral magnetic effect is expected. 
At RHIC one can select central collisions by requiring a low signal 
in the zero degree calorimeters that detect spectator neutrons. 
Then one can analyze the dependence of the signal on 
the elliptic flow present in the events. 
If the signal is due to elliptic flow one should find a direct
relations between the two. 
Calculations of the relative strength of the effect in different
scenarios are under way~\cite{Voloshin:uu2010}. 
\begin{figure}
\begin{minipage}[t]{0.45\textwidth}
\centerline{\includegraphics[width=0.9\textwidth]{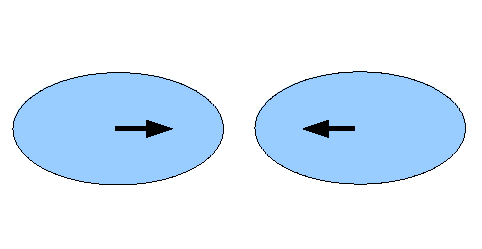} }
	\centerline{(a) }
\end{minipage}
\hspace{0.05\textwidth}
\begin{minipage}[t]{0.45\textwidth}
\centerline{\includegraphics[width=0.6\textwidth]{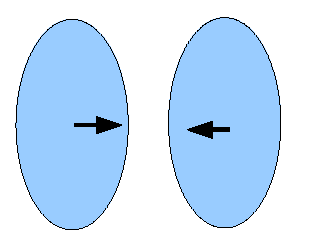} }
	\centerline{(b) }
\end{minipage}
 \caption{Schematic view of central U+U collisions:
(a) tip-tip
and 
(b) body-body.
}
\label{figUU}
\end{figure}
\item {\sl Collision of isobaric nuclei.}
The charge separation dependence on the strength of the magnetic
field can be further studied with collision of isobaric
nuclei, such as  $^{96}_{44}Ru$ and    $^{96}_{40}Zr$.
These nuclei have the same mass number, but differ by the charge.
The multi-particle production 
in the midrapidity region would be affected very
little in collision of such nuclei, and in particular one would expect 
very similar elliptic flow. 
At the same time the magnetic field would be proportional to the
nuclei charge and can vary by more than 10\%, which can results in
20\% variation in the signal.  
Such variations should be readily measurable.
The collisions of  $^{96}_{44}Ru$ and    $^{96}_{40}Zr$
 isotopes have been successfully 
used at GSI~\cite{Hong:2001tm,Hong:2003jk} in a study of baryon stopping.
Collisions of isobaric nuclei at RHIC will be also extremely valuable
for understanding the initial conditions, and in particular the
initial velocity fields, the origin of directed flow, etc.
\item {\sl Beam energy scan.} 
The charge separation effect might depend strongly on the formation 
of a quark-gluon plasma and chiral symmetry 
restoration~\cite{Kharzeev:2007jp}, 
and the signal can be greatly suppressed or completely absent
at an energy below that at which a quark-gluon plasma is formed.
Taking also into account that the life-time of the strong magnetic
field is larger at smaller collision energies, it could lead to an almost
threshold effect: with lowing the energy the signal might slowly
increase with an abrupt drop thereafter.
This questions can (and will) be addressed, for example,
during the RHIC beam energy scan.
But, unfortunately, the exact energy dependence of the chiral magnetic effect
is not calculated yet.
\item {\sl Identified particle studies.}
Large statistics recorded by RHIC experiments during the last runs(s) 
will allow 
identified and multiparticle correlations studies.
There are several questions to be addressed in such analyses.
With multiparticle correlations one can try to estimate the size 
of the cluster (the average multiplicity). 
The correlations using neutral particles should be mostly determined
by background effects, and thus provide an estimate of those.
The most interesting from my point of view, but also difficult,
would be a study of the ``quark content'' of the cluster.
The topological cluster decays in equal number of $q\bar{q}$-pairs of all
flavors, the so-called 't Hooft interaction (e.g. used in instanton
model to explain the difference in $\bar{d}/\bar{u}$ in the nucleon
``sea''). 
     
\item {\sl Clusters,  pp2pp experiment and double Pomeron collisions.}
From early FNAL and ISR measurements 
it has been known that cluster formation plays an important 
role in multiparticle production at high energies,
for a review, see~\cite{Foa:1975eu}.
These clusters, with a size of about 2--3 charged particles per cluster,
may account for production of a significant fraction of all particles.
It is interesting and important to establish how these clusters 
are related to the topological clusters as suggested 
in~\cite{Ostrovsky:2002cg} (``turning points'' -- QCD sphalerons). 
Recent progress in describing of the soft  Pomeron as a 
multi-instanton ladder~\cite{Kharzeev:2000ef} also suggests 
that the topologically nontrivial gluonic configurations play
important role played in multiparticle production. 
The properties of such clusters in principle can be studied in any
multiparticle production processes, but one possibility stands out of the list, 
namely the double Pomeron exchange~\cite{Shuryak:2003xz}, 
as illustrated in Fig.~\ref{figpp2pp}.
A specific kinematics of double Pomeron exchange process can be
addressed by pp2pp experiment~\cite{Guryn:2009zz}.
This reaction allows to investigate the properties of the 
 clusters in a very clean environment,
with one cluster per event and well controlled kinematics. 
The goal is to compare the measurements to
expectations for sphaleron decays -- invariant mass, angular
distribution of the decay products, quark composition
consistent with 't Hooft interaction, etc. 
Two pion Bose-Einstein correlation analysis will be of a particular
interest. 
If the particle production indeed is due to
decay of classical fields, the so-called chaoticity parameter $\lambda$
(intercept of the correlation function at zero relative momentum) 
could show a significant decrease.
\item{\sl CP forbidden decays.}
Not going into theoretical details, I only mention here a very
exciting possibility of an observation of \CP-forbidden decays,
e.g. $\eta \rightarrow \pi \pi$, which become possible in the presence
of \CP-odd domains~\cite{Millo:2009ar,giovanni}.  
\end{itemize}

\begin{figure}
\centerline{\includegraphics[width=0.8\textwidth]{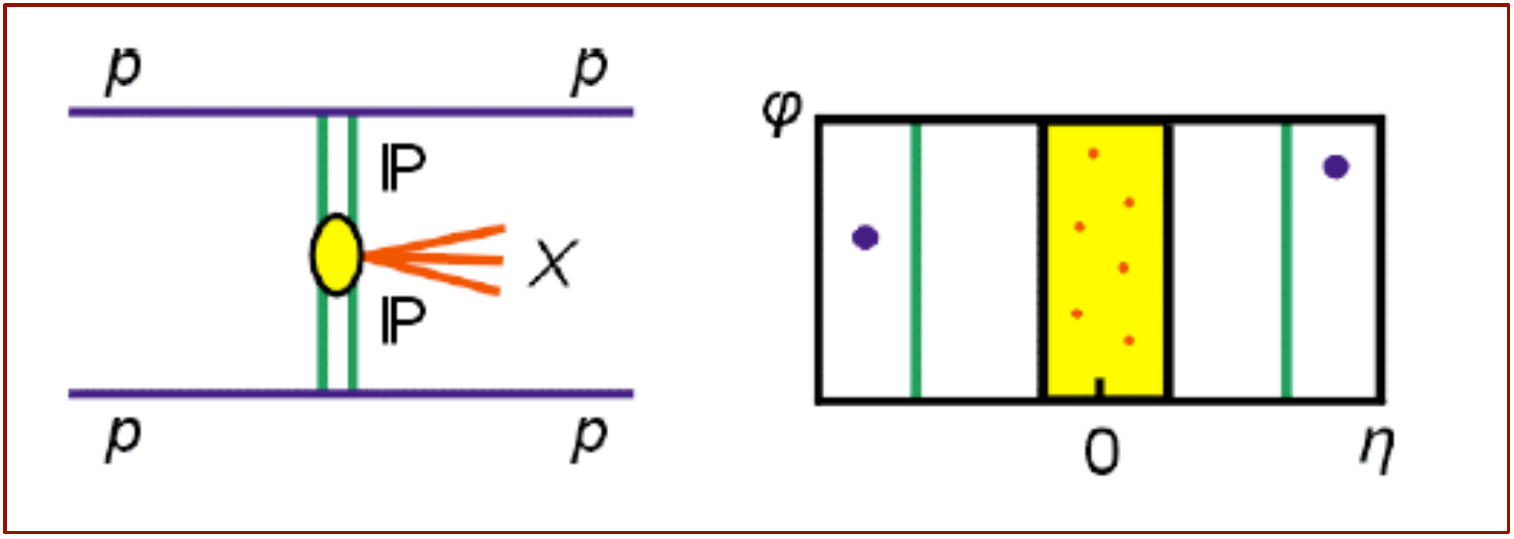} }
 \caption{Double Pomeron exchange reaction and particle distributing
   in $\eta - \phi$ plane.
}
\label{figpp2pp}
\end{figure}

\subsection{\bf \textsl{Need for a ``better'' theory.}}

For new theoretical developments, including recent lattice results, I
refer to the talk of D.~Kharzeev~\cite{KharzeevHere}. Here I only
emphasize, that 
detailed interpretation of the experimental data is not possible
without realistic theoretical calculations.
Many of needed calculations, such as dependence
on centrality and system system size, look fully doable 
though require
significant computing and man power (e.g. 3d hydrodynamics is needed for
the calculation of the magnetic field).
Detailed predictions on the transverse momentum and particle type
dependence of the effect is also essential in differentiating 
the signal from possible background contributions.
A good  theoretical understanding of the background correlations
themselves is also required, as at present all event generators lack a 
good description of correlation results.


\section{Summary.}
Experimental observation of the chiral magnetic effect
 may provide a unique opportunity
 for a direct observation of the topological structure of QCD.
The theoretical predictions are well within reach of the experiment. 
STAR has reported results that agree 
with the magnitude and gross features
of the theoretical predictions for local \P-violation in heavy-ion
collisions, but better theoretical calculations of the
expected signal and potential physics backgrounds 
are essential for further experimental study of this phenomenon. 

A very exciting future program dedicated to detail study of the effect
is emerging. It includes U+U collisions, which could serve as a test
of the chiral magnetic effect relation to the correlations
observed by STAR.

{\bf \textsl{Acknowledgments.}}  
Discussions with J.~Dunlop, D.~Kharzeev, S.~Pratt, and E.~Shuryak  
are gratefully acknowledged.

\medskip

\end{document}